\documentclass[11pt]{article}
\usepackage[dvipdfmx]{graphicx}
\usepackage[normalem]{ulem}
\usepackage{latexsym}
\usepackage{url}
\usepackage{amsmath,mathtools,amssymb}
\usepackage{pifont}
\usepackage{textcomp}
\usepackage[T1]{fontenc}
\usepackage{bbold}
\usepackage{ascmac}

\usepackage{comment}
\usepackage{cancel}

\usepackage{float}
\usepackage{multicol}
\usepackage{multirow}

\usepackage[rm,up,sc,compact,topmarks,calcwidth,pagestyles]{titlesec}

\usepackage[top=3cm,bottom=2cm,left=2.5cm,right=2.5cm]{geometry}

\def\mbf#1{\mbox{\boldmath ${#1}$}}

\newcommand{\red}[1]{\textcolor{red}{#1}}

\usepackage[rm,up,sc,compact,topmarks,calcwidth,pagestyles]{titlesec}

\titleformat{\section}[hang]{\bfseries\large}{\textbf{{\S}{\thesection}.}}{3pt}{}
\titlespacing{\section}{0pt}{18pt}{6pt}

\titleformat{\subsection}[hang]{\large}{\textbf{\thesubsection}}{3pt}{}
\titlespacing{\paragraph}{0pt}{18pt}{6pt}

\titleformat{\subsubsection}[hang]{\normalsize}{\textbf{\thesubsubsection}}{3pt}{}
\titlespacing{\paragraph}{0pt}{18pt}{6pt}

\titleformat{\paragraph}[hang]{\bfseries\normalsize}{{\theparagraph}}{1pt}{}
\titlespacing{\paragraph}{0pt}{18pt}{0pt}

\usepackage{subfigure}

\makeatletter
\newcommand{\figcaption}[1]{\def\@captype{figure}\caption{#1}}
\newcommand{\tblcaption}[1]{\def\@captype{table}\caption{#1}}
\makeatother

\usepackage{caption}
\makeatletter
  
  \@addtoreset{equation}{section}
\makeatother
\providecommand{\url}[1]{\texttt{#1}}
\providecommand{\urlprefix}{}

\providecommand{\Capitalize}[1]{\uppercase{#1}}
\providecommand{\capitalize}[1]{\expandafter\Capitalize#1}

\providecommand{\bbletal}{et~al.}

\providecommand{\bblin}{in}

\usepackage[driverfallback=dvipdfm,hyperfootnotes=false]{hyperref}
\hypersetup{
    colorlinks=true,
    citecolor=blue,
    linkcolor=red,
}
\usepackage{cite}

\usepackage{soul}
\setstcolor{red}
\newcommand{\revise}[1]{\st{#1}}
\def\mbf#1{\mbox{\boldmath ${#1}$}}

\usepackage{fancyhdr}
\begin{document}

\thispagestyle{empty}
\vspace*{3mm}
\begin{center}
  \LARGE{\textbf{Significance of Fabry-Perot cavities for space gravitational wave antenna DECIGO\\}}
  \vspace{15mm}
  \large{Kenji Tsuji$^A$, Tomohiro Ishikawa$^A$, Kurumi Umemura$^A$, Yuki Kawasaki$^A$, Shoki Iwaguchi$^A$, Ryuma Shimizu$^A$, Masaki Ando$^B$, Seiji Kawamura$^{A,\ C}$\\}
  \vspace{8mm}
  \begin{itemize}
    \setlength{\itemsep}{-1pt}
    \item[$^A$] \normalsize{\textit{Department of Physics, Nagoya University, Furo-cho, Chikusa-ku, Nagoya, Aichi 464-8602, Japan\\}}
    \item[$^B$] \normalsize{\textit{Department of Physics, University of Tokyo, Bunkyo, Tokyo 113-0033, Japan\\}}
    \item[$^C$] \normalsize{\textit{The Kobayashi-Maskawa Institute for the Origin of Particles and the Universe, Nagoya University, Nagoya, Aichi 464-8602, Japan}}
  \end{itemize}
  \vspace{8mm}
\end{center}

\begin{abstract}
  The future Japanese project for the detection of gravitational waves in space is planned as DECIGO. To achieve various scientific missions, including the verification of cosmic inflation through the detection of primordial gravitational waves as the main objective, DECIGO is designed to have high sensitivity in the frequency band from $0.1$ to $1$ Hz, with arms of length $1000$ km. Furthermore, the use of the Fabry-Perot cavity in these arms has been established for the DECIGO project. In this paper, we scrutinize the significance of the Fabry-Perot cavity for promoting this project, with a focus on the possibility of observing gravitational waves from cosmic inflation and binary compact star systems as indicators. The results show that using the Fabry-Perot cavity is extremely beneficial for detecting them, and it is anticipated to enable the opening of a new window in gravitational wave astronomy.
\end{abstract}
\vspace{6mm}

\vspace{8mm}


\section{Introduction}
The next stage of gravitational wave astronomy involves detecting gravitational waves in space. All gravitational waves observed to date, including the first detection in 2015 \cite{Abbott2016}, have been detected by ground-based detectors such as LIGO/VIRGO \cite{PhysRevX.13.041039}. These detectors have been continuously updated to conduct various scientific missions \cite{Aasi_2015,Acernese_2015}. Plans for larger-scale projects like the Einstein Telescope \cite{Punturo2010} and Cosmic Explorer \cite{Abbott2017} are underway as future generations of detectors. These detectors are designed to significantly enhance sensitivity compared to conventional ground-based detectors. Conversely, the need for extensive land to accommodate long arm lengths and the necessity to construct them deep underground{, considering the curvature of the Earth,} remain persistent issues with ground-based detectors. Furthermore, limitations in sensitivity at low frequencies, less than 10 Hz, due to Earth's seismic noise, suspension thermal noise, and Newtonian noise present an important issue in the ground-based detectors' capabilities \cite{Buikema_2020,Michimura_2022}. In response to these issues, space-based gravitational wave detectors have been planned to detect signals at low frequencies. Space-based detectors have no geographical constraints on their arm length, enabling the possibility of long baselines. Moreover, because they operate independently of the ground and do not rely on suspension systems, there is potential for significant sensitivity enhancement at low frequencies. The pioneering plan to detect gravitational waves in space is the Laser Interferometer Space Antenna (LISA) \cite{Amaro_Seoane_2012,amaroseoane2017laser}. LISA is planned to employ laser interferometry with a considerable arm length of $2.5~{\times}~10^9$ m and is specifically designed to detect gravitational waves within the milli-Hz band. However, because gravitational waves spend a considerable time within the arm, resulting in signal cancellation, the detector's sensitivity gradually diminishes above 0.1 Hz. Therefore, a sensitivity gap exists between the $0.1$ Hz band, where the LISA-type detector's sensitivity is limited, and approximately $10$~Hz, where ground-based detectors face limitations.\par
The DECi-hertz Interferometer Gravitational-Wave Observatory (DECIGO) is a future Japanese project aimed at detecting gravitational waves in space and covers the frequency range from 0.1 Hz to 10 Hz \cite{PhysRevLett.87.221103,10.1093/ptep/ptab019}. One notable feature of DECIGO in comparison to the space-based gravitational wave detector LISA is the integration of {Fabry-Perot} cavities within the arms of the interferometer. The general effect of using {Fabry-Perot} cavities in the detector is achieving higher sensitivity by enhancing the time during which photons experience the effects of gravitational waves. DECIGO also aims to improve sensitivity with a similar effect. However, in reality, when the arm length is exceedingly long, the effects of optical diffraction losses are significant, imposing limitations on the possibility of constructing cavities. Therefore, in advancing the DECIGO project, it is necessary to estimate the effects of sensitivity deterioration due to optical diffraction losses in evaluating the enhancement in sensitivity resulting from the use of {Fabry-Perot} cavities. This paper aims to quantitatively compare the sensitivity when using a Michelson interferometer versus employing the {Fabry-Perot} cavity and redefine the significance of adopting the {Fabry-Perot} cavity.\par
In Section \ref{sec:Overview_of_DECIGO}, the scientific mission within the DECIGO project and the standard design of DECIGO are outlined to provide an overview of this project. Section \ref{sec:Sensitivity} presents equations considering the effects of optical diffraction losses on the sensitivity of a Michelson interferometer and a differential Fabry-Perot interferometer. Following this, in Section \ref{sec:Beneficial_Effect}, the sensitivity to the targeted gravitational waves, namely, the primordial gravitational waves and those from binary systems, is optimized using the signal to noise ratio (SNR) as a metric, showcasing the effectiveness of the {Fabry-Perot} cavity in DECIGO. Simultaneously, the possibility of detecting these gravitational waves is discussed. Finally, Section \ref{sec:Summary} summarizes the results.

\section{Overview of DECIGO}\label{sec:Overview_of_DECIGO}

\begin{figure}[t]
  \begin{minipage}{0.49\textwidth}
    \centering
    \includegraphics[width=70mm]{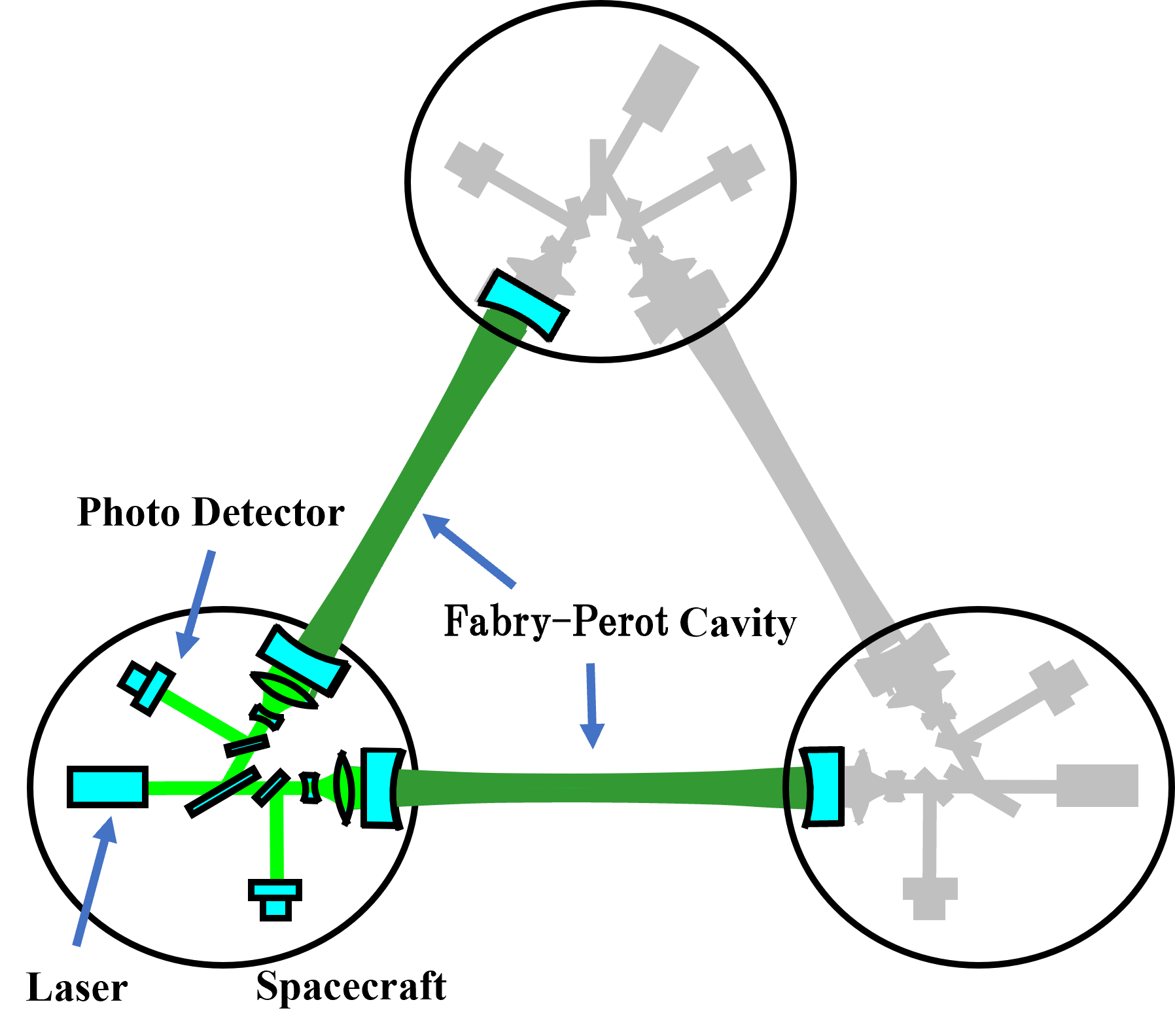}
    \caption{Configuration of a single cluster in DECIGO.}
    \label{fig:decigo1}
  \end{minipage}
  \def\@captype{table}
  \begin{minipage}[b]{0.49\textwidth}
    \begin{center}
      \tblcaption{Parameters of DECIGO's default design.}
      \label{table:Parameters of DECIGO default design}
      \begin{tabular}{lcc}
        \hline
        \hline
        Meaning       & Symbol        & \multicolumn{1}{c}{Value} \rule[0mm]{0mm}{5mm}\\[1mm]
        \hline
        Arm Length    & $L$           & $1000$ km \rule[0mm]{0mm}{5mm}\\
        Laser Power   & $P$           & $10$ W\\
        Wavelength    & $\lambda$     & $515$  nm\\
        Finesse       & $\mathcal{F}$ & $10$ \\
        Mirror Mass   & $m$           & $100$ kg\\
        Mirror Radius & $R$           & $0.5$ m\\
        \hline
        \hline
      \end{tabular}
    \end{center}
  \end{minipage}
\end{figure}

\subsection{Design of DECIGO}
\hspace{3mm} Figure \ref{fig:decigo1} shows the concept of a DECIGO cluster. Each DECIGO cluster comprises three drag-free satellites positioned in space, maintaining an equilateral triangle formation. The length of one side of the equilateral triangle, representing the distance between two satellites, is $1000$ km. Additionally, each side incorporates a Fabry-Perot cavity, forming a differential Fabry-Perot interferometer. Laser lights are installed in each satellite to compose three differential Fabry-Perot interferometers. Here, note that each cavity shares two lasers with different incident directions. Table \ref{table:Parameters of DECIGO default design} shows the parameters of DECIGO's default design. Laser power $P$ is $10$\ W, and its wavelength $\lambda$ is $515$\ nm. The mirror has a mass $m$ of $100$\ kg and a radius $R$ of $0.5$\ m. The Finesse $\mathcal{F}$, which corresponds to the effective number of light reflections within the cavity, is approximately 10. The considerably lower Finesse compared to ground-based detectors, especially in cases of long arm lengths like $1000$ km, primarily results from significant optical diffraction losses. This will be further elaborated in Section \ref{sec:Sensitivity}.\par
\begin{figure*}[h]
  \begin{minipage}[t]{0.50\textwidth}
    \centering
    \includegraphics[width=75mm]{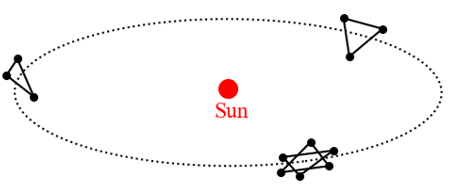}
    \caption{Positions of each cluster in DECIGO. Two clusters placed at the same position are used for correlating detection. Three clusters located at different positions have an angle of $60$ degrees to the others and are used for accurately determining the direction of gravitational waves. All clusters are in heliocentric orbits.}
    \label{fig:orbit2}
  \end{minipage}
  \begin{minipage}[t]{0.03\textwidth}

  \end{minipage}
  \begin{minipage}[t]{0.46\textwidth}
    \centering
    \includegraphics[width=75mm]{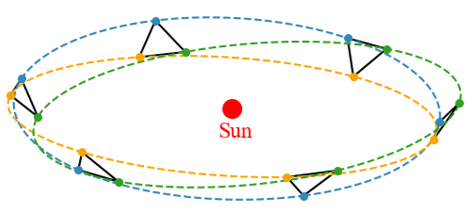}
    \caption{Proposed orbit of each cluster. All clusters are in heliocentric orbits. Each cluster maintains an equilateral triangle configuration, and the period of rotation is one year. This triangle has an inclination of $60$ degrees with respect to the plane of Earth's orbit.}
    \label{fig:orbit4}
  \end{minipage}
\end{figure*}

The entire DECIGO system consists of four clusters, each comprising satellites arranged in an equilateral triangle formation as described above, shown in Fig. \ref{fig:orbit2}. Clusters positioned in three distinct locations are utilized to enhance the angular resolution regarding the incoming direction of gravitational waves. These three locations are situated along a heliocentric orbit, also forming an equilateral triangle configuration. Furthermore, two clusters positioned in close proximity are utilized to distinctly differentiate gravitational wave signals with random incoming directions, such as primordial gravitational waves, by correlating signals between the two clusters.\par
The proposed orbit involves three satellites rotating while maintaining a triangular formation along the heliocentric orbit, as illustrated in Fig. \ref{fig:orbit4}. The triangle is inclined at an angle of 60 degrees to the plane of Earth's revolution. In general, this orbit is known as a record-disk orbit. A special case, where it is sun-synchronous and the triangle rotates once a year, is referred to as a cartwheel orbit, and it is under consideration for adoption in LISA \cite{amaroseoane2017laser,Martens_2021,JOFFRE20213868}. Since this orbit can naturally determine the motion of multiple satellites, of which the formation is composed, such as DECIGO, through a kinematic solution (Clohessy-Wiltshire equation of relative motion \cite{clohessy1960terminal}), there is a potential for minimizing modifications to the orbit.

\subsection{Science Target of DECIGO}
\hspace{3mm} Since DECIGO has higher sensitivity in the 0.1--10 Hz band compared to conventional gravitational wave detectors, numerous new scientific discoveries are anticipated.\par
Particularly, the most crucial objective of DECIGO is the observation of primordial gravitational waves. Primordial gravitational waves are assumed to have originated from quantum fluctuations of spacetime during cosmic inflation \cite{PhysRevD.37.2078,PhysRevD.42.453}. Because cosmic inflation can explain several cosmological issues, including the horizon problem and the flatness problem, it is considered the most plausible theory to explain the evolutionary process of the universe \cite{achucarro2022inflation}. The results of electromagnetic wave observations, such as those of the Planck satellite \cite{inflation2014,inflation2016,inflation2020}, are consistent with the subordinate effects caused by inflation \cite{GUTH2014112,PhysRevD.100.083537}, and further precise observations in the future are expected to strengthen the inflationary theory. However, in general electromagnetic wave observations, capturing the full extent of inflation is not possible intrinsically. This limitation arises because photons could not travel in a straight line due to interactions with electrons until approximately 380,000 years after the birth of the universe. In contrast, gravitational waves have low interaction with interfering elements, enabling them to provide insights into the universe's origins well before the 380,000 years. Therefore, the direct detection of these gravitational waves will help in determining whether inflation actually occurred, complementing conventional electromagnetic wave observations. \par
The detection of primordial gravitational waves imposes constraints on various inflation models dependent on different conditions, allowing for a detailed determination of the evolutionary processes of the early universe \cite{PhysRevD.90.063513}. Additionally, this serves as a mutual validation of observing the polarization of cosmic microwave background B-modes, which are thought to be influenced by primordial gravitational waves. Moreover, deciding the reheating temperature of the universe \cite{Kuroyanagi_2016} and testing parity symmetry in the early universe \cite{PhysRevD.75.061302} are possibilities.\par
Next, we describe the detection of gravitational waves from the coalescence of binary star systems. DECIGO excels in detecting gravitational waves from binary stars with masses on the order of $10^2M_{\odot}$ to $10^3M_{\odot}$, as detailed in Section \ref{subsec:GW_from_BSS}. The gravitational waves mentioned here refer to those generated when binary star systems approach and merge beyond the innermost stable orbit. In other words, this refers to capturing the transition from the inspiral phase to the merger phase. This typical mass is larger than the masses of binary systems observed by ground-based detectors and smaller than the masses characteristically observed by missions such as LISA. Of course, DECIGO has the ability to detect the gravitational waves from the coalescence of binary star systems with masses on the order of $10\ M_{\odot}$, which is the main objective of ground-based detectors. Gravitational waves from them persist in the $0.1$ Hz band for {several months to several years.} Due to DECIGO's lower sensitivity at high frequencies, its primary role is to serve as a precursor to ground-based detectors, providing early notifications of coalescence events. If DECIGO can detect coalescence events early, not only can ground-based gravitational wave detectors be readied to observe, but also electromagnetic and particle observatories can be readied and pointed. Conducting commonly advocated multi-messenger observations in this way makes it possible to more accurately understand information about binary systems.\par
In particular, the advancement of these multi-messenger observations is beneficial for achieving one of the important goals for DECIGO, which is to measure the cosmic acceleration expansion \cite{PhysRevLett.87.221103}. The cosmic acceleration expansion is observed as frequency shifts and phase changes. Therefore, if we can determine the redshift of the sources through electromagnetic wave observations, combining this information with gravitational wave data allows for a precise understanding of the cosmic acceleration expansion \cite{schutz1986determining,2017,Chen_2018}.\par
Furthermore, gravitational wave observations in the $0.1$ Hz range are expected to contribute to the understanding of Type Ia supernovae \cite{Maselli2020, Kinugawa_2022}. Type Ia supernovae are considered to originate from the merger of white dwarf binaries, making the detection of such merger events in the $0.1$ Hz band potentially conducive to unraveling this mechanism.\par
In addition to these, DECIGO enables the verification of general relativity \cite{Yagi_2010} and exploration of dark matter \cite{PhysRevLett.102.161101}. These aspects would contribute to a more comprehensive understanding by observing the effect of gravitational lensing on gravitational waves, which has been discussed in recent years \cite{PhysRevD.103.044005,Piorkowska-Kurpas_2021}.

\section{Sensitivity in Gravitational Wave Detectors}\label{sec:Sensitivity}
In this section, the sensitivity of gravitational wave detectors is explained. The sensitivity of gravitational wave detectors is generally limited by various sources of noise, which can be broadly categorized into three types. The first type includes noises associated with the components of the detector, such as suspension thermal noise, coating thermal noise on mirrors, and intensity noise in the laser. These noises can be significantly reduced in principle through technological advancements, such as placing mirrors in extremely low-temperature environments or not using suspension systems like space-based detectors. The second type of noise arises from the external environment, such as Earth's seismic noise and Newtonian noise. These noises can be effectively reduced by placing detectors in space. The last type of noise is quantum noise. As this noise has a quantum nature, there are special detection methods, such as homodyne detection, that formally reduce the noise. However, quantum noise is inherently inevitable. Therefore, if the other noises can be made small enough through technological advancements, quantum noise will ultimately limit the sensitivity. In this paper, {we calculate the sensitivity of the detector based on the assumption that it is limited by quantum noise.}\par
When calculating the sensitivity of a space-based detector, it is important to consider the effect of optical diffraction losses, as space-based detectors typically have very long arm lengths. In this calculation, the effect of optical diffraction losses is expressed as a decrease in laser power hitting the test mass. The beam is typically represented by Hermite--Gaussian modes (TEM$_{lm}$), which are solutions to the wave equation for the electric field obtained with the paraxial approximation. In this paper, we use the fundamental mode TEM$_{00}$ of Hermite--Gaussian modes. The intensity distribution of the TEM$_{00}$ mode has a Gaussian profile and is defined by the following \cite{svelto2010principles}.
\begin{align}\label{eq:Gaussian_beam}
    I(x,y,z) &= {I_0}\ {\exp{\left[{\frac{-2(x^2+y^2)}{w(z)^2}}\right]}}={\frac{2{P}_{\mathrm{in}}}{{\pi}w(z)^2}}{\exp{\left[{\frac{-2(x^2+y^2)}{w(z)^2}}\right]}}\\
   {{I_0}} &{=  {\frac{2P_{\mathrm{in}}}{{\pi}w(z)^2}}}\ .
\end{align}
{$I_0$ represents} 
 the radiation intensity at the center of the beam, and ${P}_{\mathrm{in}}$ denotes the total power of each beam entering each arm. However, in the Michelson interferometer, the laser power entering each arm due to the beam splitter is half of the laser power $P$ from the source. Therefore, it satisfies $P = 2{P}_{\mathrm{in}}$. $w(z)$ is the distance where the radiation intensity at the center of the beam, $I_0$, decreases to $1/e^2$. {At this time, the laser power at the end test mass located at $z$, $P_{\mathrm{ETM}}$, is given as follows:}
\begin{align}\label{eq:P_ETM}
    {P_{\mathrm{ETM}} = {\int_0^{2{\pi}}}d{\theta}{\int_0^{R_{\mathrm{ETM}}}}I(r,z)rdr = P_{\mathrm{in}}\left[1-{\exp\left(-{\frac{2R^2_{\mathrm{ETM}}}{w_0^2}}{\frac{z_R^2}{z^2+z_R^2}}\right)}\right].}
\end{align}
{{Here,}  $w_0$ represents the beam waist size. $z_R$ represents the Rayleigh range and satisfies the following equations:}
\vspace{-6pt}
\begin{align}
    {z_R} &{= {\frac{{\pi}w_0^2}{\lambda}}}\\
    {w(z)} &{= w_0{\sqrt{1+\left({\frac{z}{z_R}}\right)^2}}\ .}
\end{align}
{In addition,}  we can determine the beam waist size $w_0$ such that  {$P_{\mathrm{ETM}}$ shown in Equation \eqref{eq:P_ETM} is maximized.}
The optimal beam waist size denoted as $w_0=w_0^{\mathrm{(opt)}}$, is expressed as follows:

\begin{equation}\label{eq:w0_opt}
  w_0^{\mathrm{(opt)}} = {\sqrt{\frac{z{\lambda}}{\pi}}}.
\end{equation}
{However,} when $w_0^{\mathrm{(opt)}}$ becomes larger than the size of the test mass, it is, in practice, impossible to create a beam waist of that size. In this case, therefore, we set the beam waist size $w_0$ to be equal to the mirror radius $R_{\mathrm{ETM}}$. Summarizing the above, the power of the light incident on the end test mass ${P}_{\mathrm{ETM}}$ is given by:
\begin{equation}\label{eq:I_ETM_2}
  {P}_{\mathrm{ETM}} =
\left\{ \,
    \begin{alignedat}{2}
    &  {P}_{\mathrm{in}}\left[1-\exp{\left(-{\frac{{\pi}}{z{\lambda}}}R_{\mathrm{ETM}}^2\right)}\right]&&\hspace{3mm}\left(\text{for}\hspace{3mm} w_0^{\mathrm{(opt)}}\leq R_{\mathrm{ETM}}\right)\\
    &  {P}_{\mathrm{in}}\left[1-\exp{\left(-{\frac{2{\pi}^2R_{\mathrm{ETM}}^4}{{\lambda}^2z^2+{\pi}^2R_{\mathrm{ETM}}^4}}\right)}\right]&&\hspace{3mm}\left(\text{for}\hspace{3mm} w_0^{\mathrm{(opt)}}\geq R_{\mathrm{ETM}}\right)
    \end{alignedat}
\right.
.
\end{equation}

\subsection{Michelson Interferometer}\label{subsec:MI}
The power spectral density of noise in a Michelson interferometer $S_n^{\mathrm{(MI)}}$, which is normalized by the signal of gravitational waves {and has units of Hz$^{-1}$}, is given as follows:
\begin{equation}\label{eq:S_MI}
  \begin{split}
    S_n^{\mathrm{(MI)}} &= {{\mathcal{N}}^{\mathrm{(MI)}}_{\mathrm{RP}}}^2+{{\mathcal{N}}^{\mathrm{(MI)}}_{\mathrm{Shot}}}^2\\
    &= \left({{\kappa}^{\mathrm{(MI)}}}+{\frac{1}{{{\kappa}^{\mathrm{(MI)}}}}}\right)\frac{{h_{\mathrm{SQL}}^{\mathrm{(MI)}}}^2}{2}.
  \end{split}
\end{equation}
{The first term}  represents radiation pressure noise, and the second term represents shot noise. Moreover, ${\kappa}^{\mathrm{(MI)}}$ and $h_{\mathrm{SQL}}^{\mathrm{(MI)}}$ are given as a function of the sideband frequency $\Omega$, as~follows:
\vspace{-6pt}
\begin{align}
  {\kappa}^{\mathrm{(MI)}} &= (2A_{\mathrm{ETM}}k_0)^2{\frac{{\hbar}}{m{\Omega}^2}}\\\label{eq:h_SQL}
  h_{\mathrm{SQL}}^{\mathrm{(MI)}} &= {\frac{1}{c\ \left|{\sin{\left({\frac{L}{c}}{\Omega}\right)}}\right|}}{\sqrt{\frac{4\hbar}{m}}}.
\end{align}
{Here, $c$ is} the speed of light taken as $3{\times}10^8$ km, and $L$ represents the arm length. Furthermore, $m$ represents the mass of the mirror, and $k_0$ represents the wavenumber of the laser. $A_{\mathrm{ETM}}$ is the amplitude of the laser at the end test mass, defined using the laser angular frequency ${\omega}_0$ as follows:
\begin{equation}\label{eq:A_ETM}
  A_{\mathrm{ETM}} = {\sqrt{\frac{2{P}_{\mathrm{ETM}}}{{\hbar}{\omega}_0}}}.
\end{equation}
{In order to} express the effect of optical diffraction losses as a decrease in beam power, Equation \eqref{eq:I_ETM_2} is substituted into Equation \eqref{eq:A_ETM}, and the power spectrum is calculated. Furthermore, in this paper, we assume {the detection of interferometric signals,} using the {local} light phase-locked to the {reflected} light. This is because if the directly reflected light is adopted, laser power decreases so much that detection becomes difficult.  This implies that quantum noise is detected two times  {at both input and end ports}, and it means the power spectrum of noise in the final detection assumes a factor of 2 increase. {Thus, the power spectral density of noise in a Michelson interferometer considering the effect, ${S_n^{{\prime}\mathrm{(MI)}}}$, is given as follows:}
\begin{equation}\label{eq:S_MI2}
   {{S_n^{{\prime}\mathrm{(MI)}}} = 2{S_n^{\mathrm{(MI)}}}.}
\end{equation}
{{This method}  has also been planned for use in LISA, which is called an optical transponder \cite{PhysRevLett.123.031101,amaroseoane2017laser}.}

\subsection{Differential Fabry-Perot Interferometer}
The {power} spectral density of noise in a differential Fabry-Perot interferometer $S_n^{\mathrm{(FP)}}$ is given as follows \cite{galaxies9010009}:
\vspace{-6pt}
\begin{equation}\label{eq:S_FP}
  \begin{split}
    S_n^{\mathrm{(FP)}} &= {{\mathcal{N}}^{\mathrm{(FP)}}_{\mathrm{RP}}}^2+{{\mathcal{N}}^{\mathrm{(FP)}}_{\mathrm{Shot}}}^2\\
    {\mathcal{N}}_{\mathrm{RP}} &= {\frac{4}{mL{\Omega}^2}}{\frac{{t_{\mathrm{eff},1}}^2(r_2D)^2(1+{r_{\mathrm{eff},1}}^2)}{(1-{r_{\mathrm{eff},1}}{r_{\mathrm{eff},2}})^2}}{\sqrt{\frac{{\pi}{\hbar}P}{c{\lambda}}}}{\mathcal{H}}_{\mathrm{RP}}\\
  {\mathcal{N}}_{\mathrm{Shot}}   &= {\frac{1}{2{\pi}L}}{\frac{(1-{r_{\mathrm{eff},1}}{r_{\mathrm{eff},2}})^2}{t_{\mathrm{eff},1}(t_1D)r_{\mathrm{eff},2}}}{\sqrt{\frac{{\pi}{\hbar}c{\lambda}}{P}}}{\mathcal{H}}_{\mathrm{Shot}}\ .
  \end{split}
\end{equation}
{Here,} $r_1$ and $t_1$ represent the amplitude reflectance and transmittance of the input test mass. Similarly, $r_2$ and $t_2$ are those of the end test mass. Moreover, note that these equations already include the effect of optical diffraction losses, and the factor of its effect $D$ is defined as follows:
\begin{equation}
  D^2 = 1-{\exp\left(-{\frac{2{\pi}}{L{\lambda}}}{R^2}\right)}.
\end{equation}
{Here,} $D$ represents, for example, the already optimized $D^{\mathrm{(opt)}}$, as described in (T. \linebreak Ishikawa~et al., 2020~\cite{galaxies9010014}). Furthermore, the difference of the factor 2 in the exponent, when compared to Equation~\eqref{eq:I_ETM_2}, arises because, in our design of the differential Fabry-Perot interferometer, the position of the beam waist is the center of the cavity, defined as $z=L/2$. Using this $D$, effective amplitude reflectance ${r_{\mathrm{eff},i}}$ and the effective transmittance ${t_{\mathrm{eff},i}}$ are given as~follows:
\vspace{-6pt}
\begin{equation}
  \begin{split}
    {r_{\mathrm{eff},i}} &= r_iD^2\\
    {t_{\mathrm{eff},i}} &= t_iD^2\hspace{3mm}(i=1,2)\ .\\
  \end{split}
\end{equation}
{The factor} of $\mathcal{H}$ in each Equation \eqref{eq:S_FP} corresponds to the effect of the cavity pole, and it is given as follows:
\vspace{-6pt}
\begin{align}
  \mathcal{H}_{\mathrm{RP}}&={\frac{L{\Omega}}{c}}{\frac{1}{\left|{\sin{\left({\frac{L{\Omega}}{c}}\right)}}\right|}}\left\{1+F{\sin^2\left({\frac{L{\Omega}}{c}}\right)}\right\}^{-\frac{1}{2}}\\
  \mathcal{H}_{\mathrm{Shot}}&= {\frac{L{\Omega}}{c}}{\frac{1}{\left|{\sin{\left({\frac{L{\Omega}}{c}}\right)}}\right|}}\left\{1+F{\sin^2\left({\frac{L{\Omega}}{c}}\right)}\right\}^{\frac{1}{2}}\ .
\end{align}
{Note that}  the formula is slightly different in the referenced paper \cite{galaxies9010009} because the approximation $L{\Omega}\ {\ll}\ c$ is used for this factor in its paper. $F$ is defined by
\vspace{-6pt}
\begin{equation}
  F = {\frac{4{r_{\mathrm{eff},1}}{r_{\mathrm{eff},2}}}{(1-{r_{\mathrm{eff},1}}{r_{\mathrm{eff},2}})^2}}\ .
\end{equation}
{{In addition,}   effective finesse ${\mathcal{F}}_{\mathrm{eff}}$ is given as follows:}
\vspace{-6pt}
\begin{equation}
    {{\mathcal{F}}_{\mathrm{eff}} = {\frac{{\pi}{\sqrt{{r}_{\mathrm{eff},1}{r}_{\mathrm{eff},2}}}}{{r}_{\mathrm{eff},1}{r}_{\mathrm{eff},2}}}}.
\end{equation}
\vspace{1pt}
\subsection{Sensitivity of Gravitational Waves Detectors Cluster}

\begin{figure}[t]
  \centering
  \vspace{-3mm}
  \includegraphics[height=92mm]{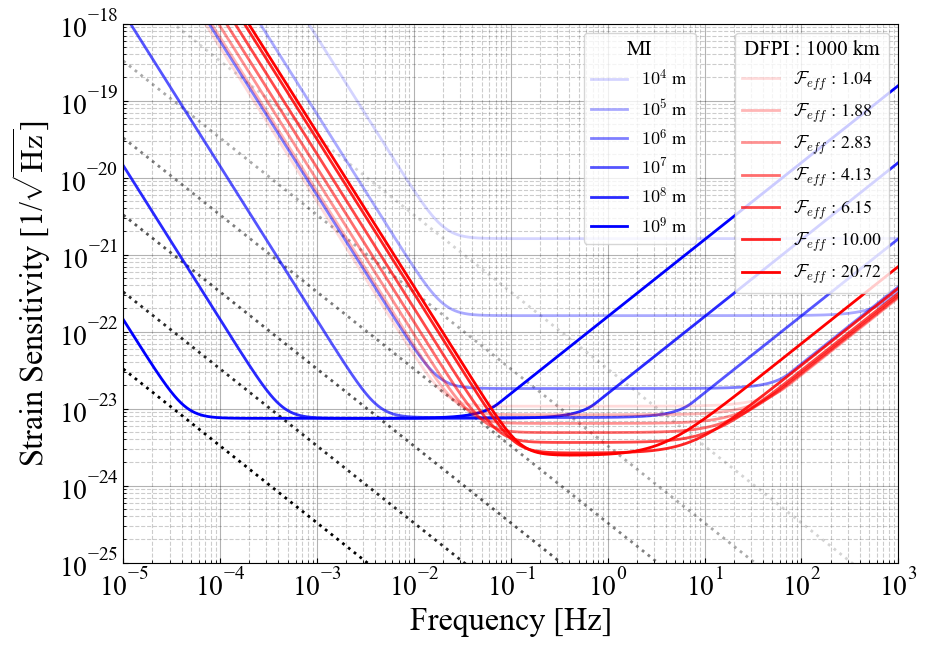}
  \caption{Typical sensitivity curves of the detectors. The Michelson interferometer (MI) is represented by the blue lines, and the color becomes lighter as the arm length shortens. The differential Fabry-Perot interferometer (DFPI) is represented by the red lines, and the color becomes lighter as the effective finesse decreases. The arm length of DFPI is fixed at $1~{\times}~10^3$~km, which is the same as the default design of DECIGO. {The dashed lines represent the standard quantum limit, as shown in Equation~\eqref{eq:h_SQL}, and the sensitivity curves of MI are not tangent to these lines due to the factor described in Section~\ref{subsec:MI}.}}
  \label{fig:Sensitivity_curve1}
  \vspace{-5mm}
\end{figure}

Space-based detectors such as DECIGO and LISA are composed of a cluster consisting of three satellites placed in an equilateral triangle and three interferometers. The sensitivity of one cluster can be considered as that of pseudo-interferometers with orthogonally intersecting two arms, integrating the sensitivities of the three interferometers \cite{PhysRevD.66.122002}. With this assumption, the power spectral density of noise for one cluster, $S^{\mathrm{cluster}}_n(f)$, is given using the power spectral of noise for a single interferometer, $S_n(f)$, as follows \cite{galaxies9010014}:
\vspace{-6pt}
\begin{equation}\label{eq:S_n_cluster}
  S^{\mathrm{cluster}}_n(f) = {\frac{2}{3}}S_n(f)\ .
\end{equation}

\subsection{Typical Noise Power Spectral Density}
To summarize this section, typical sensitivities of each detector are shown in Figure \ref{fig:Sensitivity_curve1}. The reason why the sensitivity floor level exists even when the arm length is increased in a Michelson interferometer is that the effects of power reduction due to optical diffraction losses balance with the increased capability of receiving more gravitational wave signals by extending the arm length. Focusing on this balance point, we consider shortening the arm length to the limit where the sensitivity does not deteriorate. This limitation point is understood as the point where $w_0^{\mathrm{(opt)}}=R_{\mathrm{ETM}}$ is satisfied as shown in \mbox{Equations \eqref{eq:w0_opt} and  \eqref{eq:I_ETM_2}}. If a design with $R=0.5$ m and ${\lambda} = 515$ nm like DECIGO is considered, for instance, the arm length at the limitation point is about $1.5{~\times~}10^3$ km. At this condition, using Fabry-Perot cavities in the interferometer arms is considered. The floor level of the sensitivity curve falls along about the standard quantum limit (SQL) of the Michelson interferometer when increasing the finesse of the cavity. Therefore, the sensitivity of the differential Fabry-Perot interferometer, adopted under appropriate conditions matched with the target frequency, can significantly exceed that of the Michelson interferometer, and this is the significance of using the Fabry-Perot cavity.


\section{Beneficial Effect of Employing Fabry-Perot Cavities on DECIGO}\label{sec:Beneficial_Effect}
In this section, the sensitivities of each detector are optimized for primordial gravitational waves and gravitational waves from binary systems, and the sensitivities are compared. Unless otherwise stated, the formulas in this section are referenced from (M. Maggiore, 2007 \cite{10.1093/acprof:oso/9780198570745.001.0001}).
\subsection{Primordial Gravitational Waves}\label{subsec:PGW}
\subsubsection{Wave Form}
\hspace{3mm} Various observations to date have provided limits on the amplitude of primordial gravitational waves \cite{Planck2020}, which in turn have imposed limits on the energy density of gravitational waves. By normalizing these experimental values ${\rho}_{\mathrm{GW}}$ by the critical energy density of the universe ${\rho}_c$, the upper limit of the spectrum of the primordial gravitational wave energy density ${\Omega}_{\mathrm{GW}}$ is defined as follows \cite{ mingarelli2019understanding}:
\vspace{-6pt}
\begin{equation}\label{eq:Omega_gw}
  \begin{split}
    {\Omega}_{\mathrm{GW}}(f) &= {\frac{1}{{\rho}_c}}{\frac{d{\rho}_{\mathrm{GW}}}{d\log{f}}}\\
    {\rho}_c &= {\frac{3c^2H_0^2}{8{\pi}G}}\ .
  \end{split}
\end{equation}
{Here,}  $H_0$ is the Hubble constant. Since primordial gravitational waves are expected to be observed as a superposition of waves arriving from all directions, the value that characterizes the waveform of primordial gravitational waves is not the square of the gravitational wave amplitude $h^2$ but the power spectral density of the Fourier mode, as described in (B.~S.~Sathyaprakash et al., 2009 \cite{sathyaprakash2009physics}). Hence, we also use $S_h(f)$ in this paper, and it is defined using Equation~\eqref{eq:Omega_gw} as follows:
\vspace{-6pt}
\begin{equation}\label{eq:S_h_pgw}
  \begin{split}
    S_h^{(\mathrm{PGW})}(f) = {\frac{3{H_0}^2}{4{\pi}^2}}{\frac{{\Omega}_{\mathrm{GW}}(f)}{f^3}}\ .
  \end{split}
\end{equation}
{Equation}~\eqref{eq:S_h_pgw} is defined regardless of the configuration of the detector. In contrast, for the measurement of primordial gravitational waves, to distinguish between foreground noise and the signal, correlations between two or more detectors are generally considered. The factor arising from this effect is defined as ${\gamma}(f)$ in the following:
\vspace{-6pt}
\begin{equation}\label{eq:orf}
  \begin{split}
    {\gamma}(f) &= {\frac{1}{\left<F_{+}^2\right>+\left<F_{\times}^2\right>}}\ {\Gamma}(f)\\
    {\Gamma}(f) &= {\int}{\frac{d^2\hat{\mbf{n}}}{4{\pi}}}\ {\int}{\frac{d{\psi}}{2{\pi}}}\left[{\sum_{A}}F_1^A(\hat{\mbf{n}})F_2^A(\hat{\mbf{n}})\right]\ \exp{\left(i\ 2{\pi}f{\hat{\mbf{n}}}{\cdot}{\frac{{\mbf{x_2}}-{\mbf{x_1}}}{c}}\right)},
  \end{split}
\end{equation}
where $A$ is an identifier for the mode of gravitational waves, and $A={+},{\times}$. Furthermore, $1,2$ are identifiers for the correlated detectors, and ${\mbf{x}}_{1,2}$ represents the coordinates of each detector \cite{PhysRevD.59.102002,PhysRevD.62.024004}. $F_{+},F_{\times}$ represent antenna pattern functions of detectors; their mean square, in the case of a detector type like a Michelson interferometer, is satisfied using the angle between two arms denoted as $\beta$, as follows:
\begin{equation}
  \left<F_{+}^2\right> = \left<F_{\times}^2\right> = {\frac{1}{5}}\ {\sin^2{\beta}}
\end{equation}
{Since space-based} detectors such as DECIGO and LISA generally arrange three satellites in an equilateral triangle,  we set $\beta$ to ${\pi}/3$. Moreover, in order to simplify, assuming that two detectors are located exactly at the same location, we set ${\gamma}=1$ in Equation \eqref{eq:orf}. Therefore, $\Gamma$ satisfies the following:
\begin{equation}\label{eq:Gamma}
  {\Gamma} = {\frac{2}{5}}\ {\sin^2{\beta}}\hspace{5mm}({\text{for}}\hspace{3mm}{\gamma=1})\ .
\end{equation}
\subsubsection{Signal-to-Noise Ratio}\label{subsubsec:SNR_for_PGW}
\hspace{3mm} In this paper, the signal-to-noise ratio (SNR) is used to evaluate the sensitivity to gravitational waves. For primordial gravitational waves, the SNR when correlating the signals from two detectors is defined as follows by choosing an optimized filter, as shown in (B. Allen et al., 1999 \cite{PhysRevD.59.102001}):
\begin{equation}
  \begin{split}
    \left({\frac{S}{N}}\right)^2 = 2T{\int_{f_{\mathrm{min}}}^{f_{\mathrm{max}}}}df\ {\Gamma}^2(f){\frac{S_h^2(f)}{S_{n,1}(f)\ S_{n,2}(f)}}\ .
  \end{split}
\end{equation}
{By substituting}  Equations \eqref{eq:S_h_pgw} and \eqref{eq:Gamma}, the following is obtained:
\vspace{-6pt}
\begin{equation}\label{eq:SNR_for_PGW}
  \begin{split}
      \left({\frac{S}{N}}\right)^2 &= {\frac{4}{25}}{\sin^4{\beta}}\left({\frac{3H_0^2}{4{\pi}^2}}\right)^22T{\int_{f_{\mathrm{min}}}^{f_{\mathrm{max}}}}df\ {\frac{{\Omega}_{\mathrm{GW}}^2(f)}{S_{n,1}(f)\ S_{n,2}(f)}}f^{-6}\\
      \left({\frac{S}{N}}\right) &= {\frac{3H_0^2}{10{\pi}^2}}\ {\sin^2{\beta}}\left[2T{\int_{f_{\mathrm{min}}}^{f_{\mathrm{max}}}}df\ {\frac{{\Omega}_{\mathrm{GW}}^2(f)}{S_{n,1}(f)\ S_{n,2}(f)}}f^{-6}\right]^{\frac{1}{2}}\ .
  \end{split}
\end{equation}
{Here,} $T$ represents the observation time, which has been set to a duration of $3$ years. Furthermore, the lower limit of the frequency of integral, $f_{\mathrm{min}}$, is set to $0.1$ Hz, considering the confusion limiting noise mainly caused by the white dwarf binary \cite{10.1111/j.1365-2966.2003.07176.x,galaxies10010025}. Conversely, the upper limit of the frequency, $f_{\mathrm{max}}$, is set to $1$ Hz. {The setting of this upper limit is also because the contribution to the SNR on the higher frequency side is negligible because of the strong power-law behavior of the primordial gravitational waves, and its validity is explained in detail in Section \ref{subsubsec:Result2}.} The noise power spectral densities of the detectors, $S_{n,1}(f)$, $S_{n,2}(f)$, are considered to be equal, and the values are substituted into the results calculated by Equations \eqref{eq:S_MI2},  \eqref{eq:S_FP}, and  \eqref{eq:S_n_cluster}.
\begin{table}[h]
  \begin{center}
    \caption{Parameters used to calculate the SNR of primordial gravitational waves}
    \label{table:Parameters used to calculate the SNR of PGW}
    \begin{tabular}{llc|c}
      \hline
      \hline
      Meaning                & Symbol  & \multicolumn{1}{c|}{Value} & DECIGO(Default)\rule[0mm]{0mm}{5mm}\\[1mm]
      \hline
      Laser Power            & $P$ & \multicolumn{1}{r|}{$(10,\ 30,\ 100)$ W} & $10$ W\rule[0mm]{0mm}{5mm}\\
      Mirror Radius          & $R$ & \multicolumn{1}{r|}{$(0.5,\ 0.75,\ 1)$ m} & $0.5$ m\\
      Arm Length             & $L$ & Free   & $1000$ km\\
      Amplitude Reflectance* & $r_1$ & 0 to 1 & (Finesse: $10$)\\[1mm]
      \hline
      \hline
    \end{tabular}
  \end{center}
  \mbox{}\\[1mm]
  \footnotesize
  \textbf{Note. }In the default design of DECIGO, amplitude reflectance is determined to achieve a finesse of 10.
  \normalsize
\end{table}

\begin{figure}[h]
  \centering
  \includegraphics[width=180mm]{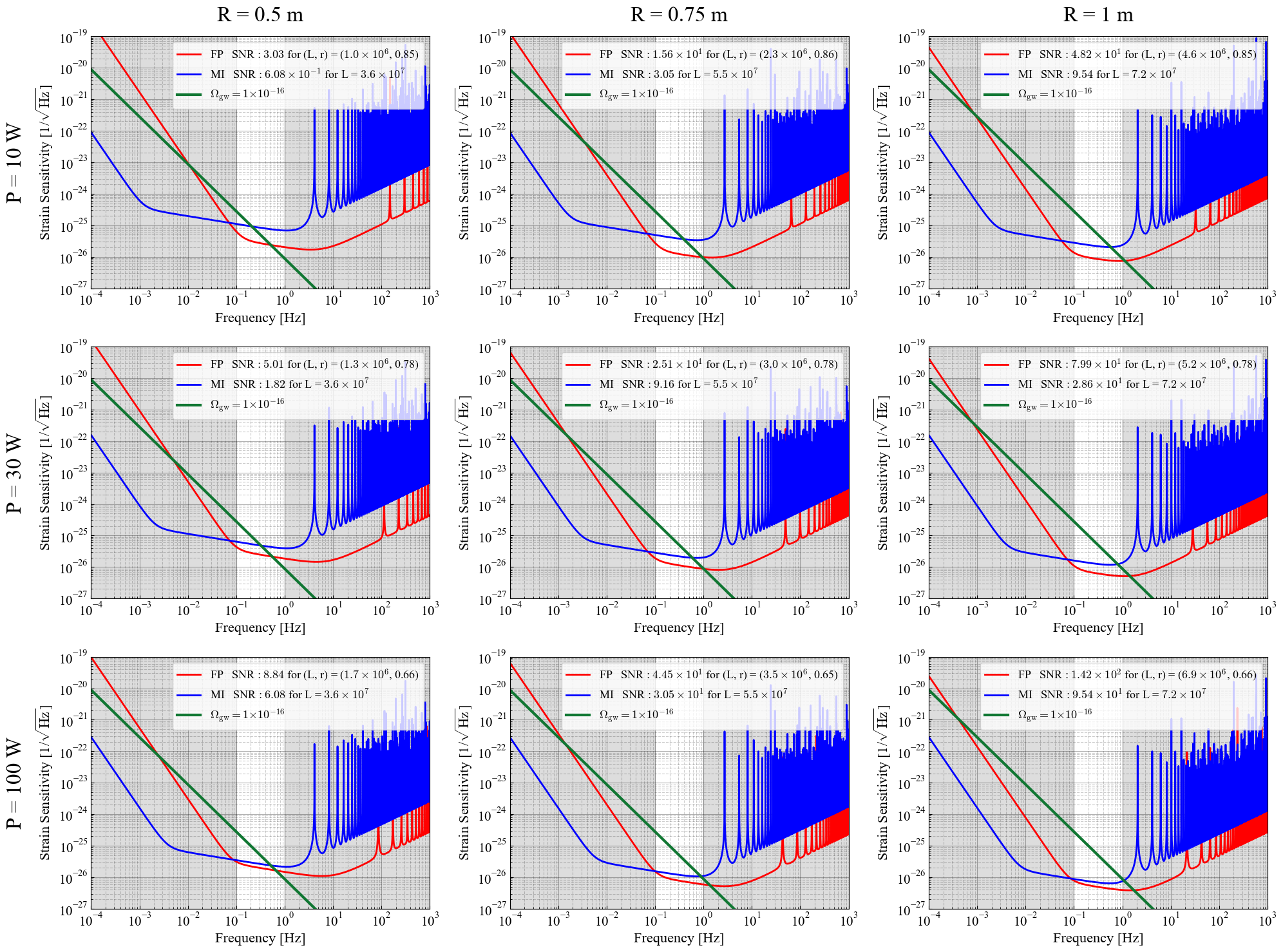}
  \caption{Comparison of detector sensitivities at an optimized SNR to detect primordial gravitational waves across different conditions. The sensitivity curves for various mirror radii are horizontally arranged, and those for different laser powers are vertically arranged, showing the optimized results for nine cases. The Michelson interferometer is represented by the blue lines, and the differential Fabry-Perot interferometer is represented by the red lines. These sensitivity curves follow Equation \eqref{eq:S_n_cluster}, with a three-year correlation. In addition, the power spectral density of primordial gravitational waves is represented by green lines, with an adopted energy density of ${\Omega}_{\mathrm{GW}}$ equal to $1~{\times}~10^{-16}$.}
  \label{fig:result_pgw_1}
\end{figure}
\clearpage
\subsubsection{Comparison of Sensitivity}
\hspace{3mm} Upon comparison of sensitivity, the parameters of each detector are optimized to maximize the SNR. Table \ref{table:Parameters used to calculate the SNR of PGW} shows parameters used as variables. Laser power $P$ and mirror radius $R$ are considered under three conditions, resulting in a total of nine patterns for sensitivity comparison. Additionally, the arm length $L$ is treated as a free parameter in the calculations. In the differential Fabry-Perot interferometer case, the amplitude reflectance of the front mirror, $r_1$, is also treated as a free parameter. Furthermore, the amplitude reflectance of the end mirror, $r_2$, is set to 1 in all cases. Moreover, the sensitivity curves $S^{\mathrm{cluster}}_n(f{,T})\left|_{\mathrm{corr}}\right.$ obtained by taking a three-year correlation are used in a figure. Using the noise power spectral density of one cluster $S^{\mathrm{cluster}}_n(f)$, this is defined as follows:
\vspace{-6pt}
\begin{equation}
  S^{\mathrm{cluster}}_n(f,T)\left|_{\mathrm{corr}}\right. = {\frac{1}{\sqrt{Tf}}}S^{\mathrm{cluster}}_n(f)\ .
\end{equation}
{Here, the} factor proportional to $1/{\sqrt{f}}$ arises from an increase in the number of observable cycles as the frequency becomes higher.


\subsubsection{Result}
\hspace{3mm} The sensitivity curves for each detector, optimized under different conditions, are illustrated in Figure \ref{fig:result_pgw_1}. These curves demonstrate that the differential Fabry-Perot interferometer consistently exhibits better sensitivity compared to the Michelson interferometer in all conditions investigated in this paper. For instance, when we focus on the DECIGO’s default design, with $P = 10$ W, $R = 0.5$ m, the differential Fabry-Perot interferometer has five times the sensitivity of the Michelson interferometer. In addition, given that the SNR of the Michelson interferometer cannot exceed an SNR of $1$ under these conditions, the incorporation of a Fabry-Perot cavity becomes a crucial factor in the detection of gravitational~waves.\par
Next, we describe the differences in detail with each condition. In general, detectors tend to achieve a higher SNR when larger mirrors are employed. This is simply because the reduction in optical diffraction losses allows more photons to be affected by gravitational waves. Similarly, enhancing the power leads to improved sensitivity, as shot noise, which is proportional to $1/\sqrt{P}$, becomes a limiting factor in the $0.1$ Hz band. However, it is evident that the effect of an increase in  power is less significant when employing the Fabry-Perot cavity compared to the Michelson interferometer. In the case of the differential Fabry-Perot interferometer, the crossover frequency is around $0.1$ Hz. Therefore, a mere increase in power causes a shift in the sensitivity curve to the lower right along the SQL without offering substantial benefits to primordial gravitational waves. To avoid these problems in the case of optimized parameters in high power, the longer arm length and the lower reflectivity to decrease the circulating power are selected. The longer arm length leads to a lower SQL, although the effect of optical diffraction losses increases. Moreover, the lower circulating power avoids a shift to the lower right along the SQL. This helps achieve a higher SNR, maintaining the crossover frequency around $0.1$ Hz. In essence, the increase in the SNR is not significant because efforts to decrease the circulating power, which increase shot noise, become necessary.\par
\begin{figure}[t]
  \centering
  \includegraphics[width=140mm]{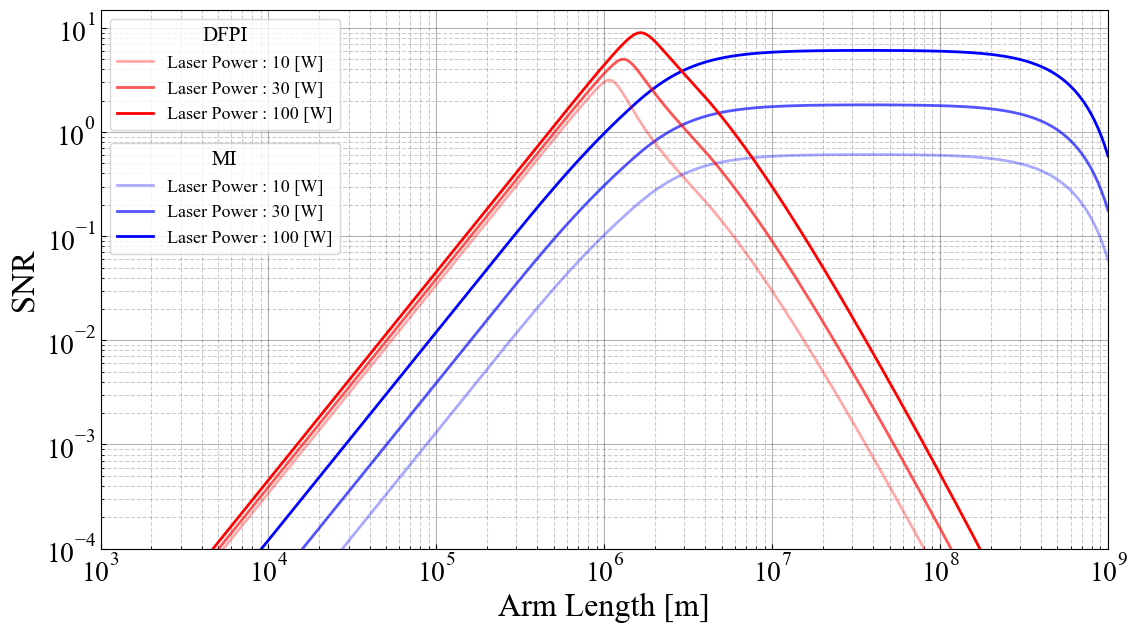}
  \caption{SNR of each detector to primordial gravitational waves, with an adopted energy density ${\Omega}_{\mathrm{gw}}$ of $10^{-16}$, for arm lengths $L$ when mirror radius $R$ is fixed at $0.5$ m. The period for taking correlations is also three years. The colors lighten as the laser power decreases in the sensitivity curve of each detector.}
  \label{fig:result_pgw_2}
\end{figure}

The change in the SNR versus arm length with the condition $R=0.5$ m is shown in Figure \ref{fig:result_pgw_2}. The distinctive feature of the differential Fabry-Perot interferometer is its ability to achieve a high SNR at lower arm lengths, although the range of length where the high SNR is attained is narrow. This is because as the arm length is extended, the increase in optical diffraction loss prevents the cavity from fulfilling its role effectively. In contrast, the Michelson interferometer achieves a high SNR over a range of arm lengths. Furthermore, especially with the adoption of long arm lengths, the SNR deteriorates, because the frequency where gravitational wave signals cancel falls within the target frequency band.

\subsection{Gravitational Waves from Coalescence of Binary Star System}\label{subsec:GW_from_BSS}
\subsubsection{Wave Form}\label{subsubsec:wave_form_bs}
\hspace{3mm} The amplitude of gravitational waves from a binary star system is given as follows:
\vspace{-6pt}
\begin{align}\label{eq:h_plus}
  {\tilde{h}}_{+}(f) &= {\frac{1}{{\pi}^{\frac{2}{3}}}}\left(\frac{5}{24}\right)^{{\frac{1}{2}}}{\frac{c}{r}}\left({\frac{GM_c}{c^3}}\right)^{\frac{5}{6}}{\frac{1}{f^{\frac{7}{6}}}}\left({\frac{1+{\cos^2{\iota}}}{2}}\right)e^{i{\Psi}_{+}(f)}\\\label{eq:h_cross}
  {\tilde{h}}_{\times}(f) &= {\frac{1}{{\pi}^{\frac{2}{3}}}}\left(\frac{5}{24}\right)^{{\frac{1}{2}}}{\frac{c}{r}}\left({\frac{GM_c}{c^3}}\right)^{\frac{5}{6}}{\frac{1}{f^{\frac{7}{6}}}}\ {\cos{\iota}}\ e^{i{\Psi}_{\times}(f)}\ .
\end{align}
Here, the chirp mass, denoted as $M_c$ and characterizing the binary star, is defined as~follows:
\vspace{-8pt}
\begin{align}
  \text{Chirp Mass : }\ & M_c = {\frac{(m_1m_2)^{\frac{3}{5}}}{(m_1+m_2)^{\frac{1}{5}}}}={\mu}^{\frac{3}{5}}{m_{\mathrm{tot}}}^{\frac{2}{5}}\\[1mm]
  \text{Reduced Mass : }\ & {\mu} = {\frac{m_1m_2}{m_1+m_2}}\\[2mm]
  \text{Total Mass : }\ & m_{\mathrm{tot}} = m_1+m_2\ ,
\end{align}
where $m_1$ and $m_2$ represent the masses of each star. Moreover, the factor ${\cos{\iota}}$ in these equations corresponds to the arrival direction of the gravitational wave. ${\Psi}_{{+},{\times}}$ represents the orbital phase and is determined by the post-Newtonian approximation \cite{PhysRevD.52.848}. The relationship between ${\Psi}_{+}$ and ${\Psi}_{\times}$ is given by the following:
\begin{equation}
  {\Psi}_{\times} = {\Psi}_{+}+{\frac{\pi}{2}}\ .
\end{equation}
{Moreover,}  the frequency $f_{\mathrm{gw}}$ of the gravitational waves emitted by a binary system is generally a time-dependent function. The time at which the binary merges is denoted as $t_{\mathrm{coal}}$, and we define ${\tau}=t_{\mathrm{coal}}-t$ using the observer's time $t$. The time evolution of the frequency is given by the following:
\vspace{-6pt}
\begin{align}
  {\dot{f}}_{\mathrm{gw}} &= {\frac{96}{5}}{\pi}^{\frac{8}{3}}\left({\frac{GM_c}{c^3}}\right)^{\frac{5}{3}}{f_{\mathrm{gw}}}^{\frac{11}{3}}\\\label{eq:f_gw}
  f_{\mathrm{gw}}({\tau}) &={\frac{1}{8{\pi}}}\left({\frac{5}{\tau}}\right)^{\frac{3}{8}}\left({\frac{GM_c}{c^3}}\right)^{-\frac{5}{8}}\ .
\end{align}
{Equation}  \eqref{eq:f_gw} indicates that while the time to coalescence $\tau$ decreases, the gravitational wave signal shifts to higher frequencies. This shift to higher frequencies reflects the fact that the distance between the binary stars, which are the sources of the waves, is decreasing. As time progresses, the two will eventually coalesce. In this context, ``coalesce'' does not refer to a simple collision between the stars; rather, it signifies that the two have reached the limit orbit where stable circular motion is possible (Innermost Stable Circular Orbit: ISCO). The typical frequency at this time, $f_{\mathrm{ISCO}}$, is defined as follows:
\vspace{-6pt}
\begin{equation}
  f_{\mathrm{ISCO}} = {\frac{1}{12{\pi}{\sqrt{6}}}}{\frac{c^3}{Gm_{\mathrm{tot}}}}\ .
\end{equation}
{Since the}  frequency $f$ here is the frequency of the wave source, the actual observation limit is $\sim$2$f_{\mathrm{ISCO}}$\ Hz.
\subsubsection{Signal-to-Noise Ratio}\label{subsubsec:SNR_for_BS}
\hspace{3mm} The amplitude of  gravitational waves $h(t)$ actually obtained by the detector is defined using the amplitudes $h_{+}$ and $h_{\times}$ for the two modes, and the antenna pattern functions $F_{+}$ and $F_{\times}$ for the detector, as~follows:
\begin{equation}
  h(t) = h_{+}F_{+}+h_{\times}F_{\times}\ .
\end{equation}
{The Fourier } transform ${\tilde{h}}$ is given by
\begin{equation}\label{eq:h}
  \begin{split}
    {\tilde{h}}(f) &= {\frac{1}{{\pi}^{\frac{2}{3}}}}\left(\frac{5}{24}\right)^{{\frac{1}{2}}}{\frac{c}{d_L}}\left({\frac{G\mathcal{M}_c}{c^3}}\right)^{\frac{5}{6}}{\frac{1}{f^{\frac{7}{6}}}}\left[F_{+}\left({\frac{1+{\cos^2{\iota}}}{2}}\right)+iF_{\times}{\cos{\iota}}\right]e^{i{\Psi}_{{+}}(f)}\ .
  \end{split}
\end{equation}
{Here, note}  that redshifted chirp mass $\mathcal{M}_c$ and luminosity distance $d_L$ are used, instead of the general chirp mass $M_c$ and normal coordinate distance $r$ in Equations \eqref{eq:h_plus} and \eqref{eq:h_cross}, in order to reflect the effect of cosmological redshift. Taking the average over the arrival direction and polarization of the gravitational waves, Equation \eqref{eq:h} can be expressed as~follows:
\vspace{-6pt}
\begin{equation}\label{eq:h_revised}
  \begin{split}
    {\tilde{h}}(f) &= {\frac{1}{{\pi}^{\frac{2}{3}}}}\left(\frac{5}{24}\right)^{{\frac{1}{2}}}{\frac{c}{d_L}}\left({\frac{G\mathcal{M}_c}{c^3}}\right)^{\frac{5}{6}}{\frac{1}{f^{\frac{7}{6}}}}\left<\left|F_{+}\left({\frac{1+{\cos^2{\iota}}}{2}}\right)+iF_{\times}{\cos{\iota}}\right|^2\right>^{\frac{1}{2}}e^{i{\Psi}_{{+}}(f)}\\
    &= {\frac{1}{{\pi}^{\frac{2}{3}}}}\left(\frac{5}{24}\right)^{{\frac{1}{2}}}{\frac{c}{d_L}}\left({\frac{G\mathcal{M}_c}{c^3}}\right)^{\frac{5}{6}}{\frac{1}{f^{\frac{7}{6}}}}\left({\frac{2}{5}}{\sin{\beta}}\right)e^{i{\Psi}_{{+}}(f)}\ .
  \end{split}
\end{equation}
Here, $\beta$ represents the angle between the two arms of the Michelson interferometer. The SNR of the detector is given by the noise power spectrum of the detector $S_n(f)$ as follows:
\begin{equation}
  \left({\frac{S}{N}}\right)^2 = {\frac{5}{6}}{\frac{1}{{\pi}^{\frac{4}{3}}}}{\frac{c^2}{d_L^2}}\left({\frac{G\mathcal{M}_c}{c^3}}\right)^{\frac{5}{3}}{\frac{4}{25}}{\frac{3}{4}}{\int_{f_{\mathrm{min}}}^{f_{\mathrm{max}}}}df{\frac{f^{-\frac{7}{3}}}{S_n(f)}}\ .
\end{equation}
{The lower}  limit of the frequency, as described in Section \ref{subsubsec:SNR_for_PGW}, is set at $0.1$ Hz, taking into account the influence of confusion limiting noise. 
Furthermore, from the discussion in Section~\ref{subsubsec:wave_form_bs}, the upper limit of the frequency satisfies $f_{\mathrm{max}}=2f_{\mathrm{ISCO}}$. Combining these factors, the SNR is defined as follows:
\begin{equation}
  \left({\frac{S}{N}}\right)^2 = {\frac{1}{10}}{\frac{1}{{\pi}^{\frac{4}{3}}}}{\frac{c^2}{d_L^2}}\left({\frac{G\mathcal{M}_c}{c^3}}\right)^{\frac{5}{3}}{\int_{0.1}^{2f_{\mathrm{ISCO}}}}\ df\ {\frac{f^{-\frac{7}{3}}}{S_n(f)}}\ .
\end{equation}
\subsubsection{Comparison of Sensitivity}
The parameters used for optimization are the same as those for primordial gravitational waves and are listed in Table \ref{table:Parameters used to calculate the SNR of PGW}. Moreover, since the dimensions of the sensitivity ${\sqrt{S_n(f)}}$ of each detector and the Fourier mode of the gravitational wave amplitude $\tilde{h}(f)$ are generally different, an equivalent quantity with aligned dimensions as ${\sqrt{S_h(f)}}$ is introduced as follows \cite{Moore_2015}:
\begin{equation}\label{eq:equivalent_S_h}
  \sqrt{S_h(f)} = f^{\frac{1}{2}}\left|{\tilde{h}}(f)\right|\ .
\end{equation}
These equivalent quantities have dimensions of $1/{\sqrt{\mathrm{Hz}}}$. When illustrating the spectrum of the gravitational wave amplitude from binary star systems, $\sqrt{S_h(f)}$ is used.

\subsubsection{Results}
\begin{figure}[H]
  \centering
  \includegraphics[width=180mm]{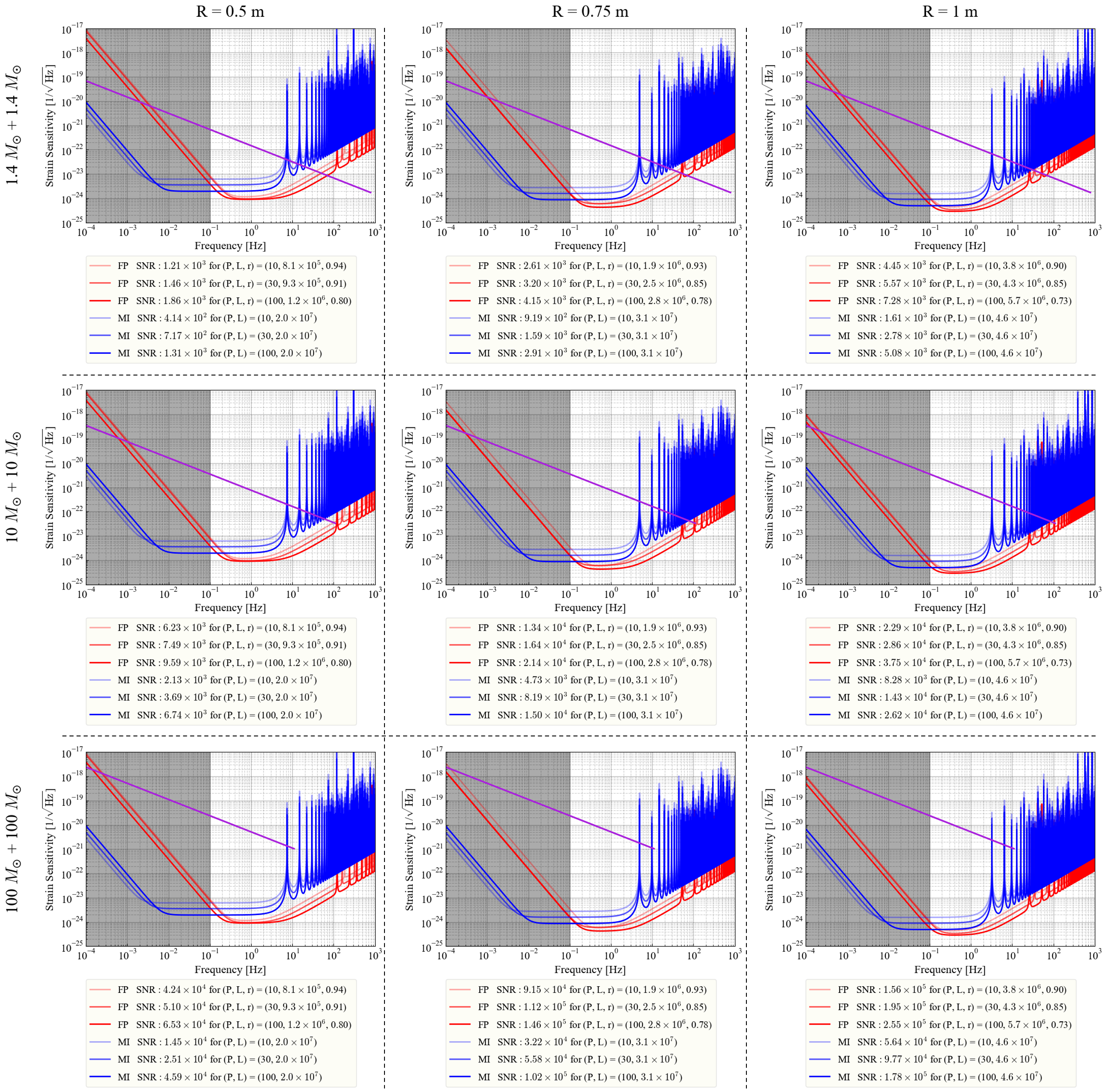}
  \caption{Comparison of detector sensitivities at an optimized SNR for the detection of gravitational waves from binary star systems across different conditions. The sensitivity curves for various mirror radii are horizontally arranged, and those for different source masses are vertically arranged, showing the optimized results for nine cases. The differences in laser power are represented by the varying shades of color. The purple line represents ${\sqrt{S_h}}(f)$ for each mass condition at $100$ Mpc from the detectors, which is equivalent to the strain sensitivity of the detector as indicated by Equation \eqref{eq:equivalent_S_h}.}
  \label{fig:result_bs_1}
\end{figure}

\hspace{3mm} The results, optimized under various conditions for gravitational waves from binary stars, are shown in Figure \ref{fig:result_bs_1}. These results indicate that the differential Fabry-Perot interferometer can achieve a higher SNR than the Michelson interferometer in all cases for gravitational waves from binary stars, similar to the case of primordial gravitational waves. Once again, the cases of DECIGO's default design with a mirror radius of $R = 0.5$\ m and laser power of $P = 10$\ W are focused on. In these cases, the SNR of the differential Fabry-Perot interferometer is $2$ to $3$ times that of the Michelson interferometer. Generally, since binary star systems are assumed to be distributed approximately isotropically in space, increasing sensitivity by a factor of 2 to 3 results in an 8 to 27 times increase in observable gravitational wave events from binary stars for the detector. Therefore, the use of the Fabry-Perot cavity provides effective advantages in these observations.\par

Next, a detailed explanation of the sensitivity curves for each condition is provided. The basic trend is similar to that for primordial gravitational waves; the sensitivity of each detector improves with higher power and a larger mirror radius. In contrast, a difference with primordial gravitational waves is that there are cases where the crossover frequency has not reached $0.1$ Hz. The frequency dependence of the spectrum of gravitational waves from binary star systems is lower than that of primordial gravitational waves. Therefore, depending on the conditions, achieving a deep floor level on the high-frequency side is more advantageous than setting a shallow floor level at $0.1$ Hz. When using the Fabry-Perot cavity, increasing the amplitude reflectance of front mirrors, i.e., increasing finesse, allows for higher internal power and enables the achievement of a lower floor level.\par

\begin{figure}[t]
  \centering
  \includegraphics[width=130mm]{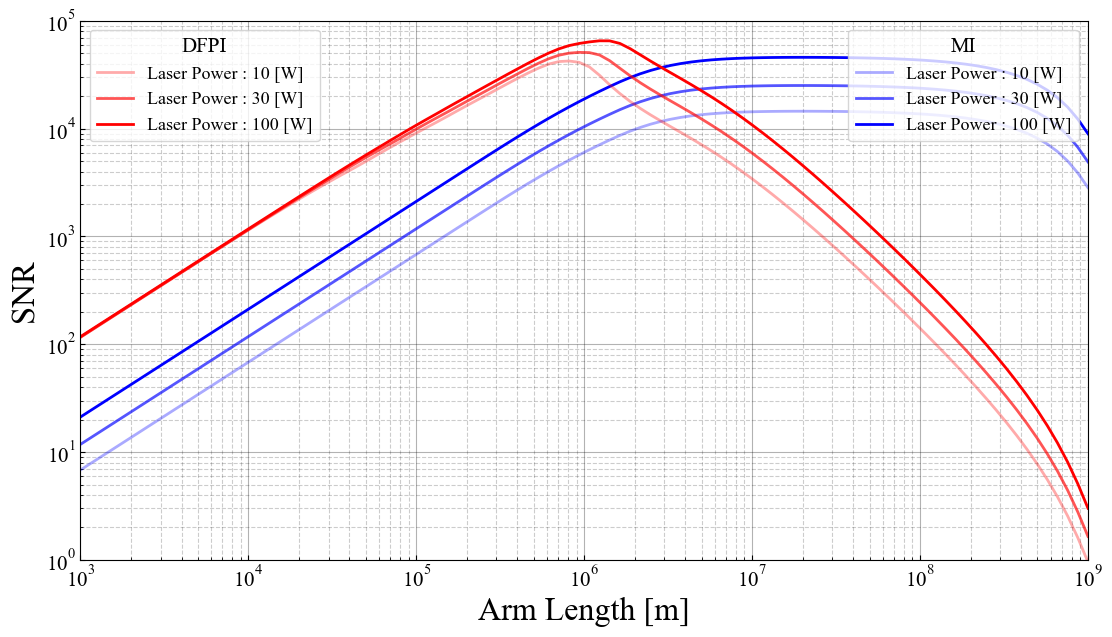}
  \caption{SNR of each detector to gravitational waves from binary star systems for arm lengths $L$ when mirror radius $R$ is fixed at $0.5$ m.  Here, the mass of the binary star systems is fixed at $100M_{\odot} + 100 M_{\odot}$. The distance from the detector of the binary system is also set to 100 Mpc. The colors lighten as the laser power decreases in the sensitivity curve of each detector.}
  \label{fig:result_bs_2}
\end{figure}

The relationship between the arm length and SNR for the default design of DECIGO with $R=0.5$ m is shown in Figure \ref{fig:result_bs_2}. The trend here is also similar to that for primordial gravitational waves, with a peak around $10^6$ m for the differential Fabry-Perot interferometer. For the Michelson interferometer, the highest sensitivity continues in the range of $10^7$ m to $10^9$ m. When comparing conditions with equal power, the results indicate that the effect of the differential Fabry-Perot interferometer is more significant for lower power.\par

\begin{figure}[t]
  \centering
  \includegraphics[width=140mm]{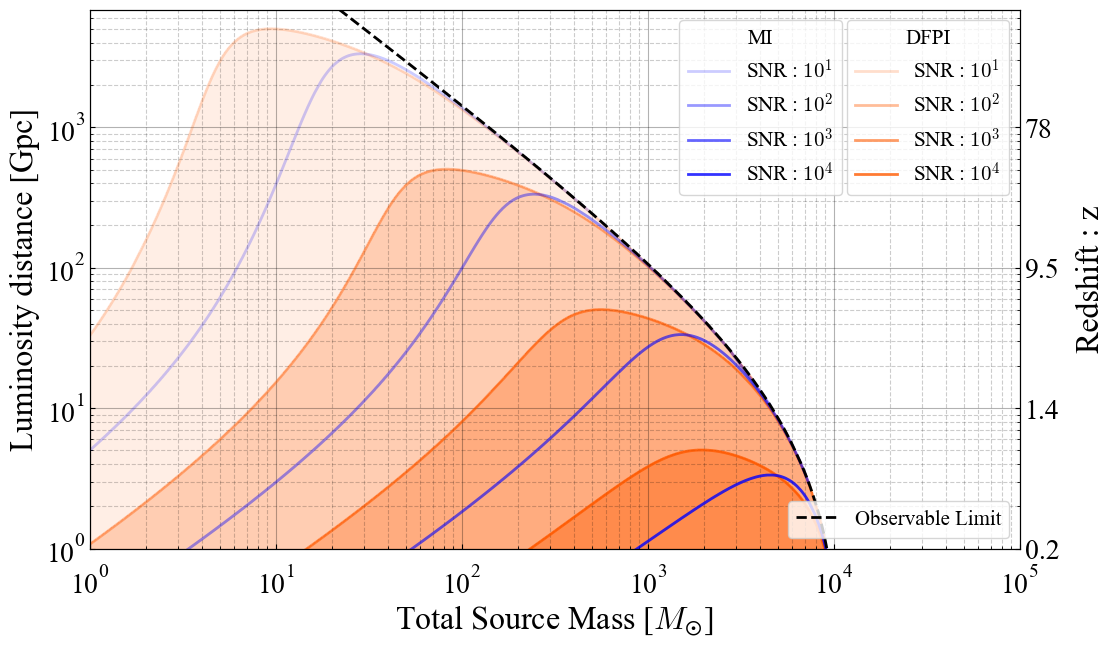}
  \caption{Relationship between the total source mass and luminosity distance of observable binary systems for each detector at various SNRs. The parameters of the detector are set as $R=0.5$ m and $P = 10$ W. The black line represents the observable limit where the redshifted merger frequency is equal to the cutoff frequency related to confusion limiting noise.}
  \label{fig:result_bs_3}
\end{figure}

Finally, the observable range of each detector is discussed. The relationship between the total source mass and luminosity distance for each detector at various SNRs is shown in Figure \ref{fig:result_bs_3} in the case of a mirror radius of $R = 0.5$ m and laser power of $P = 10$ W. Considering the observable range with an SNR greater than 10, the differential Fabry-Perot interferometer covers almost the same observable range as the Michelson interferometer, with the capability of capturing gravitational waves from low-mass binary systems located extremely far from the detector. In contrast, the observable distances on the high-mass side are almost the same. The observational limit on the high-mass side exists because the frequency at coalescence is lower than the cutoff frequency due to confusion limiting noise. For the primary target of DECIGO in the observation of gravitational waves from binary stars, which is in the range of $10^2M_{\odot}$ to $10^3M_{\odot}$, the differential Fabry-Perot interferometer is always dominant in terms of the SNR. In this mass range, for a typical value corresponding to a redshift $z=10$, approximately meaning $d_L=100$ Gpc, it consistently achieves an SNR of the order of 100.

\clearpage
\section{Summary and Prospects}\label{sec:Summary}
DECIGO is a gravitational wave detector designed to fill the observable frequency gap between ground-based detectors and low-frequency detectors like LISA. It targets gravitational waves in the frequency range from $0.1$ to $1$ Hz and has various scientific missions, including the verification of inflation theory through the observation of primordial gravitational waves. In this paper, the use of the Fabry-Perot cavity, which is the most notable feature of DECIGO, was focused on, and the differences in sensitivity of detectors, whether they have cavities or not, to gravitational waves from cosmic inflation and binary stars were discussed.\par
Regarding primordial gravitational waves, attaining higher sensitivity was confirmed when using the Fabry-Perot cavity by adjusting the arm length and mirror reflectivity, regardless of laser power and mirror radius. In the default design of DECIGO, a five-fold sensitivity improvement in the SNR was achieved, and sensitivity well above 1 in the SNR was obtained, by using the Fabry-Perot cavity. The sensitivity difference between the two detectors was observed to narrow with the increase in laser power. However, this is crucial in addressing technical challenges and has the potential to reduce the requirement of laser power in the design. Moreover, for the observation of primordial gravitational waves, advanced methods to achieve high sensitivity, such as the use of quantum locking with an optical spring, have been considered \cite{YAMADA2020126626, YAMADA2021127365, PhysRevD.107.022007,galaxies11060111}. Continuously developing and effectively combining these techniques is a current and future challenge to enhance the observability of primordial gravitational waves.\par
Similarly, sensitivity was generally better when using the Fabry-Perot cavity for gravitational waves from binary stars. In the default design of DECIGO, an improvement in the SNR by 2 to 3 times was confirmed. If binary systems are distributed isotropically in space, the number of observable gravitational wave events scales with the cube of the observable distance, and the effect is extremely significant. Additionally, achieving high sensitivity not only allows the standalone operation of DECIGO but also opens up possibilities for combining it with ground-based gravitational wave detectors or electromagnetic wave detectors. This combination is expected to dramatically improve the accuracy of parameter determination for binary star systems.\par
In summary, these results demonstrate the utility of the Fabry-Perot cavity in DECIGO and raise significant expectations for the observation of various types of gravitational waves. Through gravitational wave observations with DECIGO equipped with the Fabry-Perot cavity, we anticipate profound insights into various science targets such as the processes of cosmic formation and measurements related to the accelerated expansion of the universe.\\[5mm]
\noindent
\large
\textbf{Author Contributions}\\
\normalsize
Conceptualization, K.T. and S.K.; methodology, K.T., T.I, Y.K, S.I. and S.K.; software, K.T.; validation, T.I., M.A., and S.K.; formal analysis, T.I. and M.A.; investigation, K.T.; resources, S.K.; data curation, S.K.; writing-original draft preparation, K.T.; writing-review and editing, all; visualization, K.T. and S.K.; supervision, S.K.; project administration, S.K.; funding acquisition, S.K.\\[5mm]
\noindent
\large
\textbf{Funding}\\
\normalsize
This work was supported by JSPS KAKENHI, Grants No. JP22H01247.\\[5mm]
\noindent
\large
\textbf{Data Availability Statement}\\
\normalsize
There is no experimental data for paper.\\[5mm]
\noindent
\large
\clearpage
\noindent
\textbf{Acknowledgements}\\
\normalsize
We would like to thank Naoki Seto for helpful discussion. We also would like to thank David H. Shoemaker for commenting on a draft.\\[5mm]
\noindent
\large
\textbf{Conflicts of Interest}\\
\normalsize
 The authors declare no conflict of interest.



\end{document}